\documentclass[aps,pra,superscriptaddress,showpacs,showkeys,amsfonts,amssymb,amsmath]{revtex4}

\usepackage{epsfig}

\newcommand{\be}{\begin{eqnarray}}
\newcommand{\ee}{\end{eqnarray}}

\newcommand{\n}{\nonumber}

\newcommand{\bra}[1]{\langle#1|}
\newcommand{\ket}[1]{|#1\rangle}

\begin{document}


\title{Localization and Recurrence of  Quantum Walk in Periodic Potential on a Line}


\author{C.-I. Chou}
\affiliation{{Department of Physics, Chinese Culture University, Taipei 111, Taiwan, China }}

\author{C.-L. Ho
\footnote{email: hcl@mail.tku.edu.tw}}
\affiliation
{Department of Physics, Tamkang University,
 Tamsui 251, Taiwan, China}

\begin{abstract}

We present  numerical study of a model of quantum walk in periodic potential on the line.  We take the simple view that different potentials affect differently the way the coin state of the walker is changed.  For simplicity and definiteness, we assume the walker's coin state is unaffected at sites without potential, and is rotated in an unbiased way according to
Hadamard matrix at sites with potential. This is the simplest and most natural model of a quantum walk in a periodic potential with two coins.
Six generic cases of such quantum walks were studied numerically.  It is found that of the six cases,  four cases display significant localization effect, where the walker is confined in the neighborhood of the origin for sufficiently long times. Associated with such localization effect is the recurrence of the probability of the walker returning to the neighborhood of the origin.   

\end{abstract}



%

 \maketitle




\section{Introduction}


Quantum walks are the quantum analogues of the classical random walks (for review, see e.g.: \cite{Kempe,VA}). 
It was originally proposed with the aim of finding 
quantum algorithms that are faster than classical algorithms for the same problem.  
There are two distinct types of quantum walks, namely, discrete time quantum walks with a quantum coin  on the line \cite{ADZ,NV,LZG0,XZ,QX} and on graphs \cite{AAKV,WZTQW},
 and continuous time quantum walks \cite{FG,RSLLZG}.
 Some new quantum algorithms  based on quantum walks have been proposed. For instance, 
 a quantum search algorithm based on  discrete time quantum walk architecture 
 has been shown to gain an algorithmic speedup over classical algorithms \cite{SKW}, and a continuous time quantum walk was shown to be able to find
 its way across a special type of graph exponentially faster than any classical algorithms \cite{CCDFGS}.
 
 That quantum walks can escalate many classical algorithms lies in the fact that  in general quantum walks 
 diffuse faster than its classical counter parts.  
 For a process that gives a  symmetric distribution of the walker's positions, the tendency of diffusion can be 
 measured by the standard deviation of the position $\sigma (t)$  as a  function of time (step)  $t$.  For  classical random walk ,
  one has $\sigma (t)\propto \sqrt{t}$, but for a unbiased quantum walk on a line with a Hadamard coin (so-called Hadamard walk), one has $\sigma (t)\propto t$.  
  Analytical results  for quantum walk limit distributions have since been established \cite{Konno1}. 
 
 On the other hand, in \cite{RMM} it was shown that a quantum random walk  in a one-dimensional chain 
 using several types of biased quantum coins, arranged in aperiodic sequences following the Fibonacci prescription, 
 can lead to a sub-ballistic wave-function spreading.  
 A model with multiple internal coin states was shown to exhibit localization of the walker at the origin in \cite{IK}.
 In \cite{SK}  it was shown that  for a class of inhomogeneous quantum walks with multiple coins periodic in position, which is a generalization of the model introduced in \cite{LS},
 there could be localization at the origin for certain choices of the parameters defining the model.  Furthermore, they have shown, through numerical studies, 
 that the eigenvalue spectrum 
 of such inhomogeneous walks could exhibit a fractal structure similar to that of the Hofstadter butterfly.   Konno has presented 
 and proved a theorem on return probability for a type of space-inhomogeneous walks \cite{Konno2}.  Localization is 
 also observed in a quantum walk with  two coins operating at different times \cite{Machida}.
 Later it was found that changing a phase (i.e., imposing discontinuity) at a point in a discrete quantum walk results 
 in certain localizationn effect  \cite{Wojcik}. 
 In \cite{Shikano1} the differences in limit distributions between the classical random walks and a few models of quantum walks  were presented.
 For a recent review on these issues, see e.g., Ref. \cite{Shikano2}.
 
 The above studies indicate that suitable modifications of the position and/or coin space could lead to a rich variety of possible 
 wave function evolutions of the quantum walker.
  
In this work we would like to study if localization of quantum walker could happen in a periodic potential, i.e.,
 if the coins are arranged in a periodic way in space.  It tuns out that this is possible in the model we present here. 

To the best of our  knowledge, a model of quantum walk in a periodic potential on a line was first considered in \cite{LZG}.  There it was assumed that at sites without potential 
the walker walks as normal (coined)  Hadamard walk, but when there is potential the walker walks according to the scattering quantum walks introduced in  \cite{Scatter}. 
 Localization phenomenon was not reported in this work.

It has been shown recently that the scattering walk is equivalent to the coined walk  \cite{VAL}. Thus it is not necessary to invoke
two different approaches to describe the quantum walk in a periodic potential.  
Here we shall adopt the discrete coined quantum walks at positions throughout the line.  We take the natural assumption that the
presence of potential only affect the way the coin is flipped.  Hence we assume two different coins: one for positions 
without potential, and the other for positions with potential.  We find that for certain potentials and initial states, 
localization and recurrence of the walkers around the origin are possible.

The plan of the paper is as follows.  In Sect. II we define the model of quantum walk in a periodic potential on the line. Six generic cases of such quantum walks are then discussed with numerical results in Sect. III.  Sect. IV concludes the paper.

%
%

\section{The model}

The  discrete-time, or coined quantum walk on  a line is defined as follows \cite{ADZ,AAKV,Kempe}.
The total Hilbert space is given
by $\mathcal{H}\equiv\mathcal{H}_{P}\otimes\mathcal{H}_{C}$, where
$\mathcal{H}_{P}$ is spanned by the orthonormal vectors $\{\ket{x}, x=0,\pm 1, \pm 2, \ldots\} $
representing the positions of the walker, and $\mathcal{H}_{C}$
is a two-dimensional coin space spanned by two orthonormal vectors
denoted by $\ket{0}$ and $\ket{1}$. The dynamics of the walk is controlled by a coin flip operator $C$, which modifies the coin states of the walker, and a
conditional shift operator $S$ that shifts the walker's position according to the latest state of the coin.  Thus the evaluation operator for one step of walk is
$U=S\cdot (C\otimes I)$.   If the initial state of the walker and the coin  is $\ket{\psi_0}$,  then after $t$ steps of the walk the state of the system is $\ket{\psi (t)}=U^t\ket{\psi_0}$. 
In the original discrete quantum walk proposed in \cite{ADZ,NV} , the coin operator $C$ is taken to be the Hadamard matrix
\begin{eqnarray}
H=
 \frac{1}{\sqrt{2}}
 \left( \begin{array}{cc}
  1 & 1 \\
  1 & -1
  \end{array}\right),\label{H}
\end{eqnarray}
and the position displacement operator is given by
\begin{equation}
S=\ket{0}\bra{0}\otimes \sum_x \ket{x+1}\bra{x} + \ket{1}\bra{1}\otimes \sum_x\ket{x-1}\bra{x}.
\end{equation}
It is known that  the  initial state of the walker and coin given by
\begin{equation}
\ket{\psi_0}=\frac{1}{\sqrt{2}}\left(\ket{0} + i \ket{1}\right)\ket{0}
\end{equation}
gives rise to an outgoing symmetric probability distribution
on the positions when a single Hadamard coin is used.
On the other hand, the state
\begin{equation}
\ket{\psi_0}=\frac{1}{\sqrt{2}}\left(\ket{0} +  \ket{1}\right)\ket{0}
\end{equation}
leads to an asymmetric distribution. 

To  define a model of quantum walk in  a periodic potential on a line, we take the simple view that different potentials affect differently the way the coin state of the walker is changed.  Thus we suppose the coin state is changed by a coin operator $C_0$ when there is no potential,  and by $C_p$ when the field is present.  The 
position displacement operator at each position is
\begin{equation}
S_x=\ket{0}\bra{0}\otimes \ket{x+1}\bra{x} + \ket{1}\bra{1}\otimes \ket{x-1}\bra{x}.
\end{equation}
Together the evolution operator is
\be
U=\sum_{x: {\rm no\  potential} }S_x\left(C_0\otimes I \right)  + \sum_{x: {\rm at\  potential} }S_x\left(C_p\otimes I \right). 
%
\ee
The most general two dimensional unitary coin operator is given by
\be
C=\left( \begin{array}{cc}
  \sqrt{\rho} & \sqrt{1-\rho}e^{i\theta}  \\
  \sqrt{1-\rho}e^{i\phi} & -\sqrt{\rho}e^{i(\theta + \phi)}  \end{array}
  \right),
  \label{gencoin}
 \ee
where $0\leq \theta, \phi \leq \pi$ are arbitrary angles, and $0\leq \rho \leq 1$.
By choosing the parameters $\rho, \theta$ and $\phi$, one can assign different coin operators $C_0$ and $C_p$ to design 
a quantum walk model in periodic potential on a line.

The state of the walker after $t$ steps is
\be
\ket{\psi(t)}=U^t \ket{\psi_0}
=\sum_{x=-\infty}^\infty \left[A_x(t) \ket{0} + B_x(t) \ket{1}\right]\,\ket{x},
\ee
where $\ket{\psi_0}$ is the initial state, and
\be
\sum_x |A_x(t)|^2 + |B_x(t)|^2 =1.
\ee
At time $t$ the probability that the walker is to be found at position $\ket{x}$ is
\be
p(x,t)= |A_x(t)|^2 + |B_x(t)|^2.
\ee

%
%

\section{Numerical results}

In this work, for simplicity and definiteness, we shall choose $C_0=I$ and $C_p=H$. These two choices correspond, respectively,  to 
$\{\rho=1,\theta+\phi=\pi \}$ and $\{\rho=1/2, \theta=\phi=0\}$ in (\ref{gencoin}).
This means that the walker's coin state is unaffected at sites without potential, and is rotated in an unbiased way according to the
Hadamard matrix at sites with potential. This is the simplest and most natural model of a quantum walk in a periodic potential with two coins.

We shall be interested in symmetric walks such that $\langle x(t) \rangle=0$.  So the main physical quantity to characterize the quantum walk is the standard 
deviation $\sigma(t)=\sqrt{\langle x(t)^2 \rangle}$.  Also, as we have in mind the possibility of localization of the walker at the origin, we shall also consider the probability $P_0(t)$ of the walker at the origin $x=0$ as a function of the step $t$.

Below we shall consider six generic cases of quantum walk in periodic potential on a line with a period $N$.  
For simplicity, we shall adopt the notation $[C_1:q, C_2: (N-q)]$ to denote the situation where the coin operator $C_1$ is to be used in the first $q$ positions  and the coin $C_2$ is used in the remaining $N-q$ positions.  The origin $x=0$ is always assumed to be at the middle-point of the first $q$ positions (so in this work $q$ is always taken to be odd). 

The six cases are:
\be
{\rm Case\  IA}:   [H:1,~~I: N-1],   & ({\rm\ use\  coin}\  H \ {\rm at}\  x\equiv 0\  ({\rm mod}\ N)\  {\rm and}\  I\  {\rm elsewhere}),  &N=2,3,4,\ldots ;\n\\
{\rm Case\  IB}:   [I:1,~~H: N-1],  & ({\rm\ use\  coin}\  I \ {\rm at}\  x\equiv 0\  ({\rm mod}\ N)\  {\rm and}\  H\  {\rm elsewhere}),  &N=2,3,4,\ldots ;\n\\
{\rm Case\  IIA}:   [H:N-1,~~I: 1], & ({\rm\ use\  coin}\  I \ {\rm at}\  x\equiv N/2\  ({\rm mod}\ N)\  {\rm and}\  H\  {\rm elsewhere}), &N=2,4,6,\ldots ;\n\\
{\rm Case\  IIB}:   [I:N-1,~~H: 1],  & ({\rm\ use\  coin}\  H \ {\rm at}\  x\equiv N/2\  ({\rm mod}\ N)\  {\rm and}\  I\  {\rm elsewhere}), & N=2,4,6,\ldots; \n\\
{\rm Case\  IIIA}:   [H: q,~~I: q],  & ({\rm\ use\  coin}\  H \ {\rm at}\  x< (q+1)/2  ({\rm mod}\ 2q)\  {\rm and}\  I\  {\rm elsewhere}),  & q=3,5,7,\ldots; \n\\
{\rm Case\  IIIB}:   [I: q,~~H: q],  & ({\rm\ use\  coin}\  I \ {\rm at}\  x< (q+1)/2  ({\rm mod}\ 2q)\  {\rm and}\  H\  {\rm elsewhere}), & q=3,5,7,\ldots; \n\\
\ee

\begin{figure} 
 \centering
 \includegraphics{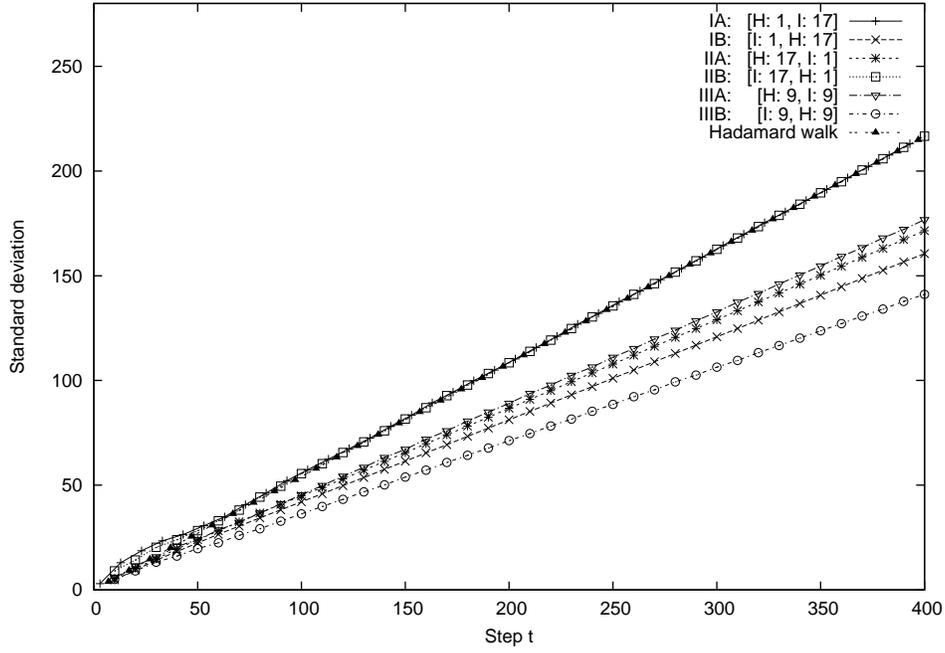}
\caption{The standard deviation $\sigma (t)$ versus the number of step $t$. 
Note that the graphs of $\sigma(t)$ for quantum walks of case IA and IIB overlap with that for the Hadamard walk at large $t$. 
For the other cases, the slope of the $\sigma(t)$ are smaller.   This means that diffusion in cases other than IA and IIB are slower, with case IIIB being the slowest.}
\label{fig1}
\end{figure}

In Fig.~\ref{fig1} we show the standard deviation $\sigma (t)$ versus the number of step $t$.  It is seen that, as in the standard unbiased 
Hadamard quantum walk, $\sigma (t)$ is generally asymptotically linear in $t$, i.e. $\sigma(t)\propto t$.

 It is also noted that $\sigma (t)$ 
for cases IA and IIB overlap with that of the standard Hadamard walk at large steps.  The other cases have smaller slopes, with that for case IIIB being the smallest. 
This means that diffusion in cases other than IA and IIB is slower than that of the standard Hadamard walk.

\begin{figure}
 \centering
 \includegraphics{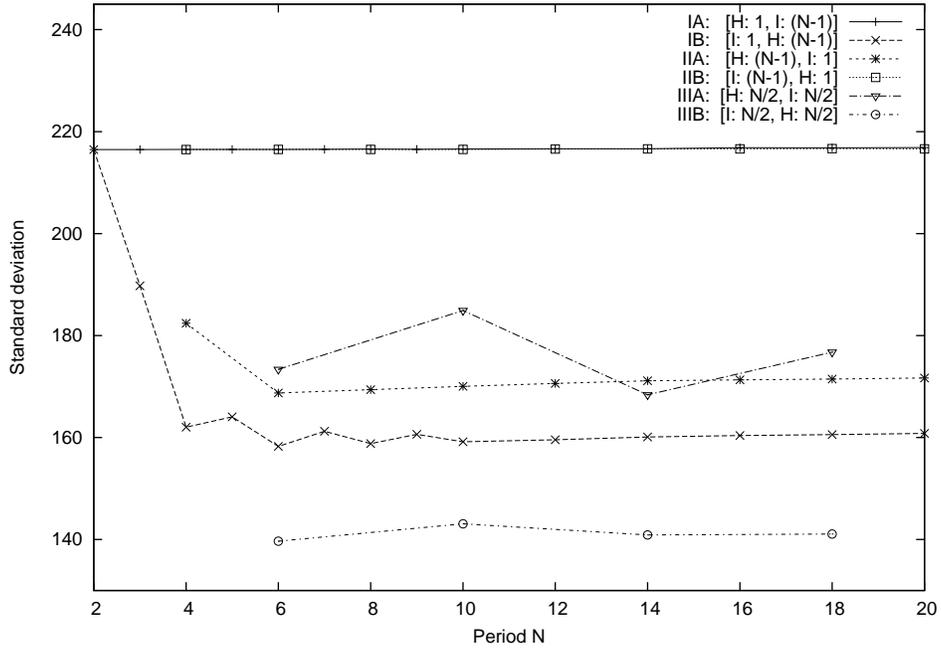}
\caption{Values of $\sigma (t)$ at a fixed step $t=400$ for different period $N$.
For cases IB and IIA, there exists critical value of the period $N$ at which the value of $\sigma (t)$ drops by a significant amount.}
\label{fig2}
\end{figure}

That such is the case can be further seen  from the values of $\sigma (t)$ at a fixed step (say, $t=400$) for different period $N$, as depicted in Fig.~\ref{fig2}.
Furthermore, for cases IB and IIA, there exists critical value of the period $N$ at which the value of $\sigma (t)$ drops by a significant amount.  To further understand the decrease in the value of $\sigma (t)$, we shall consider each case in more detail below.

%
%

\subsection{Cases IA and IB}

In these cases, either the identity coin or the Hadamard coin is used at the position $\ket{x}$  such that $x\equiv 0 ~~({\rm mod}\ N)$.

\begin{figure}
 \centering
 \includegraphics{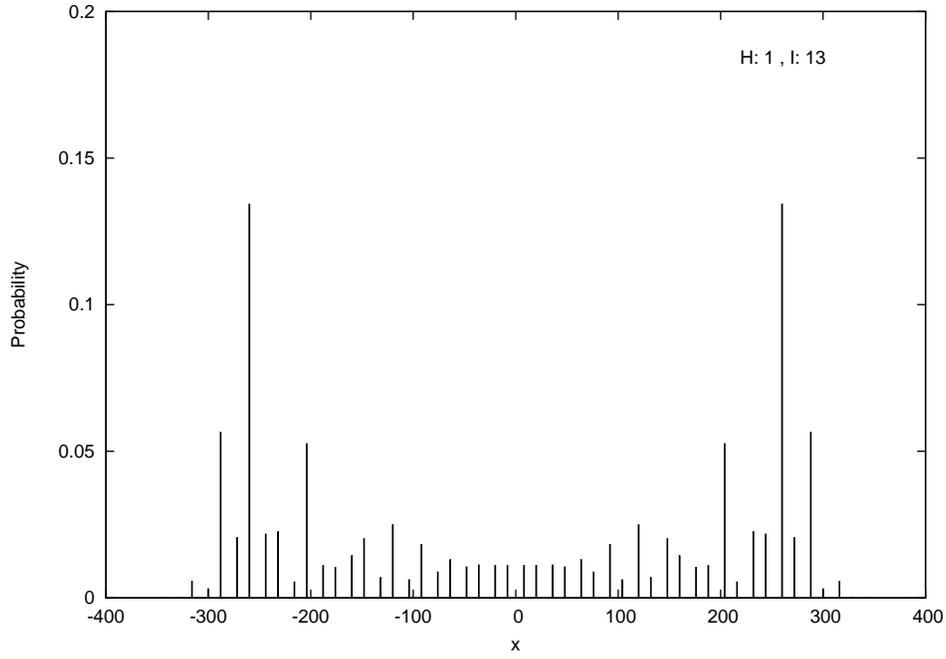}
\caption{Probability $P(x,t)$ of the quantum walk of case IA ( $[H:1, I:13]$)  with period $N=14$ at  $t=400$. Note that the probability at positions with odd $x$ are zero. There is no localization.}\label{fig3}
\end{figure}

\begin{figure}
 \centering
 \includegraphics{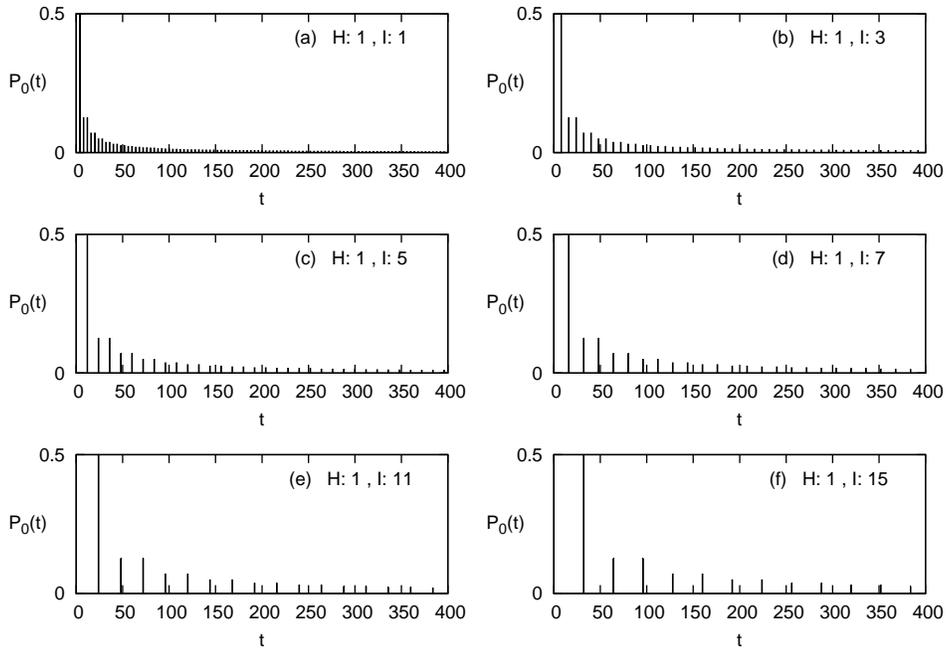}
\caption{Probability $P_0(t)$ at the origin of the quantum walk of  case IA.
The number of steps is taken up to 400.  Note that the probability at the origin must be zero when the step t is odd. This is the same for the all other cases considered in this work. There is no localization as $P_0(t)$ dies away at large steps.}
\label{fig4}
\end{figure}

For case IA, the walker encounters potential field only at the positions $\ket{x}$ such that $x$ is a multiple of the period $N$.
It is found that two peaks symmetric with respect to the origin just move away from the origin as in the case of the Hadamard walk.  Fig.~\ref{fig3} shows the probability $P(x,t)$ at $t=400$ steps.    In Fig.~\ref{fig4} we show the probability $P_0(t)$ of the walker appearing at the origin $x=0$. It is seen that in the probability just die out gradually.

\begin{figure}
 \centering
 \includegraphics{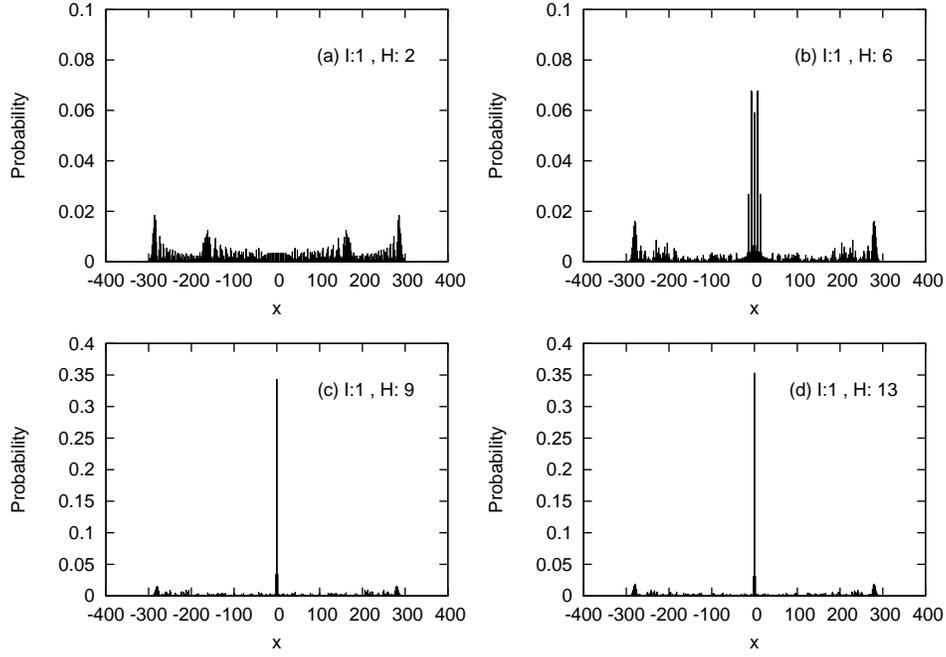}
\caption{ Probability $P(x,t)$ of the quantum walk of case IB ( $[I: 1, H: N-1]$)  with period $N=3,7,10$ and $14$ at  $t=400$. Localization at the origin occurs for $N>3$.}\label{fig5}
\end{figure}

\begin{figure}
 \centering
 \includegraphics{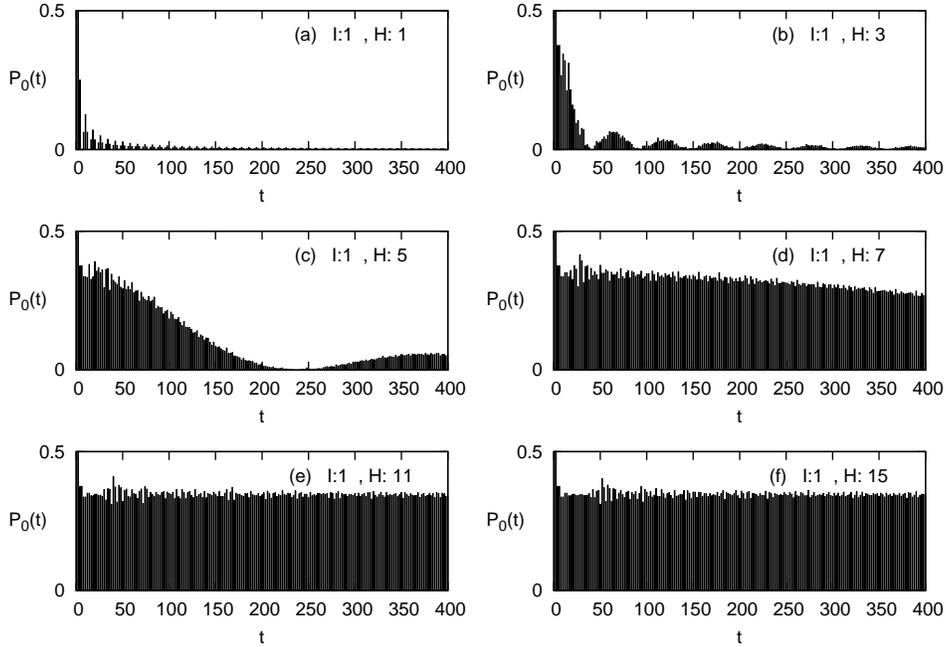}
\caption{ Probability $P_0(t)$ at the origin of the quantum walk of  case IB.
The number of steps is taken up to 400.  Recurrence of probability at the origin occurs for $N>3$. The number of steps for recurrence becomes larger as $N$ increases, indicating a stronger localization effect. }
\label{fig6}
\end{figure}

Case IB is the dual of the previous case: positions with and without potential are exchanged. 
For $N\leq 3$, the system diffuses in a similar pattern as the ordinary quantum walk with a single Hadamard coin: two symmetric peaks moving away from the origin. 
But for $N> 3 $, there appears high probability that the walker is in the neighborhood of the origin $x=0$: localization of the walker is possible.  Fig.~\ref{fig5} shows the situation for 
 four different periods $N$.  Fig.~\ref{fig6} shows the probability $P_0(t)$ that the walker returns to the origin.  For $N=2$ the probability is small as $t$ increases. But for $N=4$, we see slight recurrent increase in $P_0(t)$ as a function of $t$: recurrence of walker at the origin appears.   The tendency is stronger as $N$ increases. 

%
%

\subsection{Cses IIA and IIB}

In the cases discussed in section A, the origin at which the  walker starts walking is the only point whose force field is different from the other $(N-1)$ points in a single period $N$.  The cases considered here represent a shifted situations: the origin is at the center of the $(N-1)$ points with the same force field. This requires $N$ to be even.

\begin{figure}
 \centering
 \includegraphics{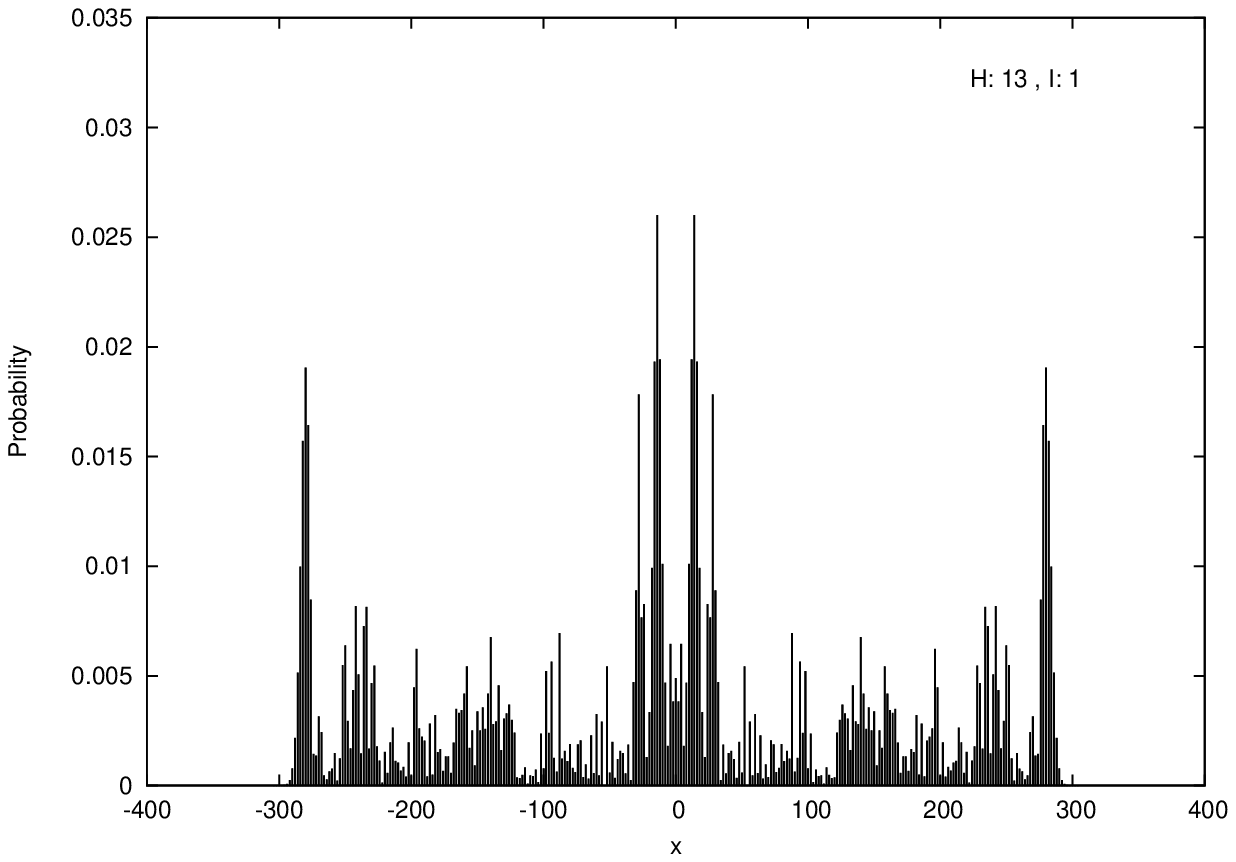}
\caption{Probability $P(x,t)$ of the quantum walk of case IIA ( $[H:13, I:1]$)  with period $N=14$ at  $t=400$. Localization at the origin occurs for $N>5$.}\label{fig7}
\end{figure}

\begin{figure}
 \centering
 \includegraphics{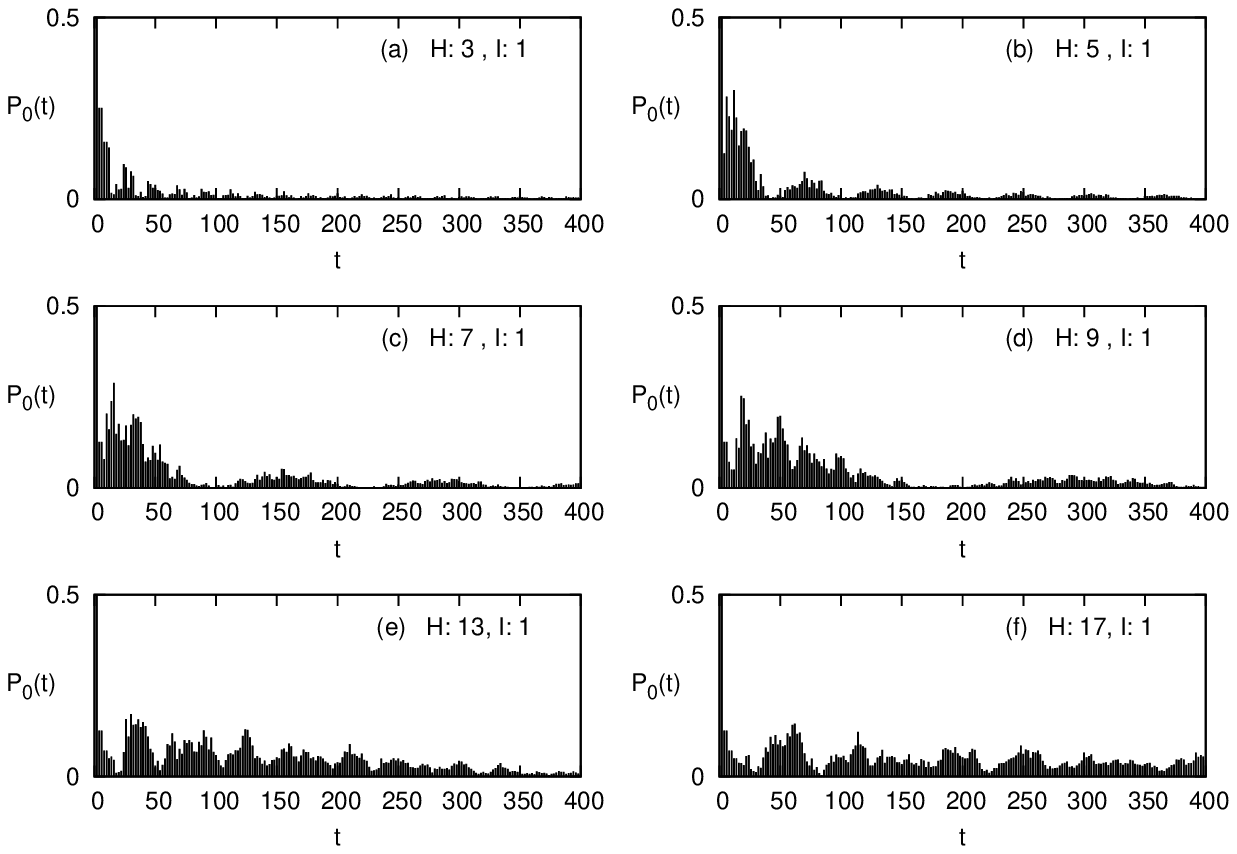}
\caption{Probability $P_0(t)$ at the origin of the quantum walk of  case IIA.
The number of steps is taken up to 400.  Significant recurrence occurs for $N>5$.}\label{fig8}
\end{figure}

In case IIA,  the walker begins its walk at the middle of the potential, and encounters no field at the positions $x$ such that $x\equiv N/2\  ({\rm mod}\  N)$. 
For $N\leq 5$, the probability distribution is symmetric with two major peaks moving away from the origin. But as $N$ increases the rate of diffusion becomes slower.  
When $N> 5$, localization  near the origin appears.  Fig.~\ref{fig7} show the situation for $N=14$.  Recurrence with high probability at the origin also appears accordingly
as depicted in Fig.~\ref{fig8}.

\begin{figure}
 \centering
 \includegraphics{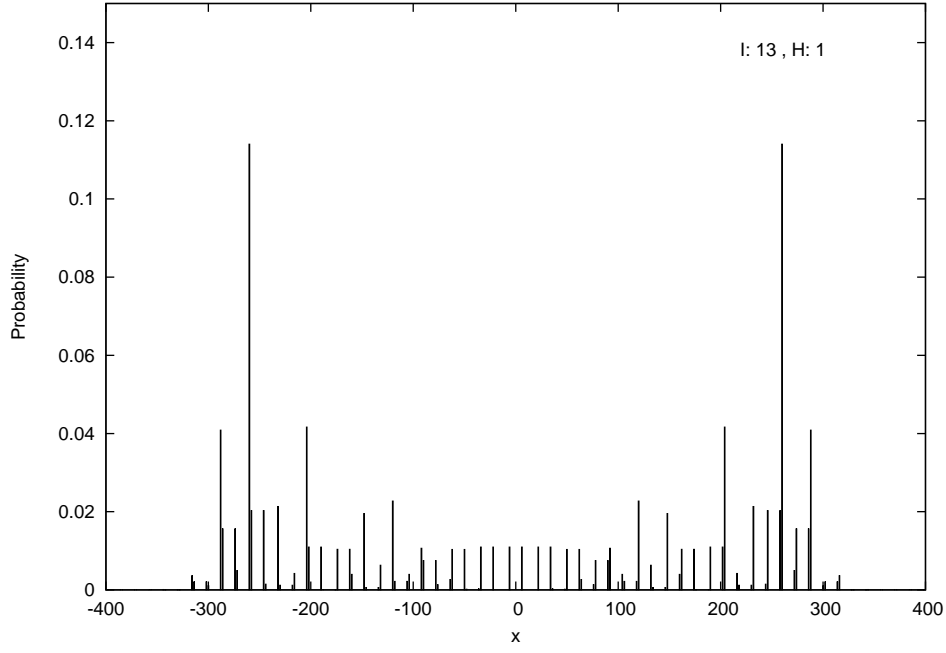}
\caption{ Probability $P(x,t)$ of the quantum walk of case IIB ( $[I:13, H:1]$)  with period $N=14$ at  $t=400$. There is no localization.}\label{fig9}
\end{figure}

\begin{figure}
 \centering
 \includegraphics{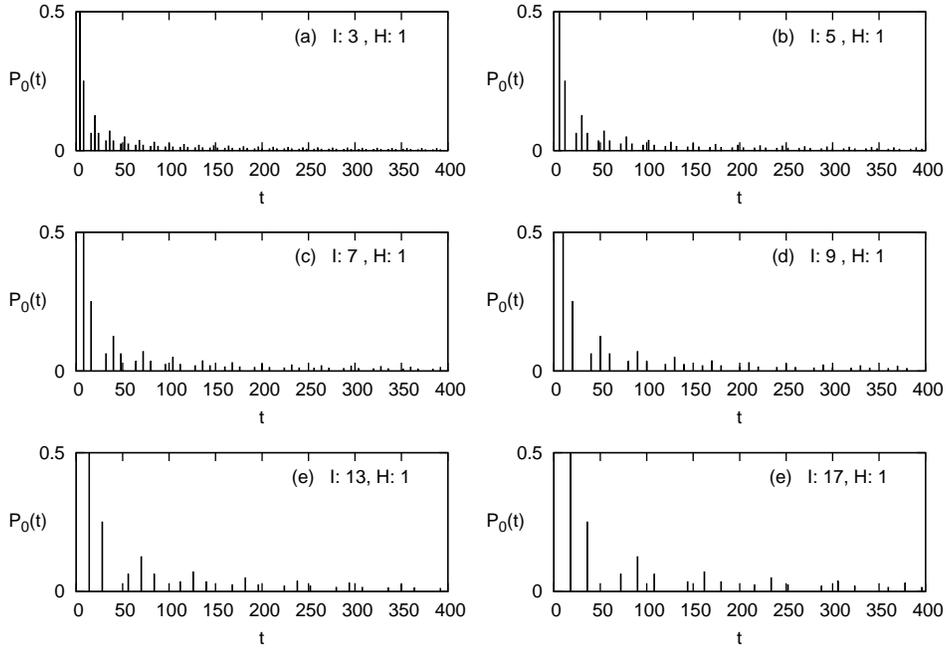}
\caption{ Probability $P_0(t)$ at the origin of the quantum walk of  case IIB.
The number of steps is taken up to 400. There is no recurrence.}\label{fig10}
\end{figure}

Again, case IIB is  the dual of the case IIA:  the walker begins its walk at the middle of no potential region, and encounters field at the 
positions $x$ such that $x\equiv N/2\  (mod\  N)$.
From Figs.~\ref{fig9} and \ref{fig10}, it is seen that there is no localization and recurrence.

%
%

\subsection{Case IIIA and IIIB}

These cases correspond to periodic potential (period $N=2q$) where locations with and without potential have equal length  $q$.
In case IIIA,  the walker begins its walk at the middle of the potential, while in case IIIB the walker begins its walk at the middle of zero  potential region.

\begin{figure}
 \centering
 \includegraphics{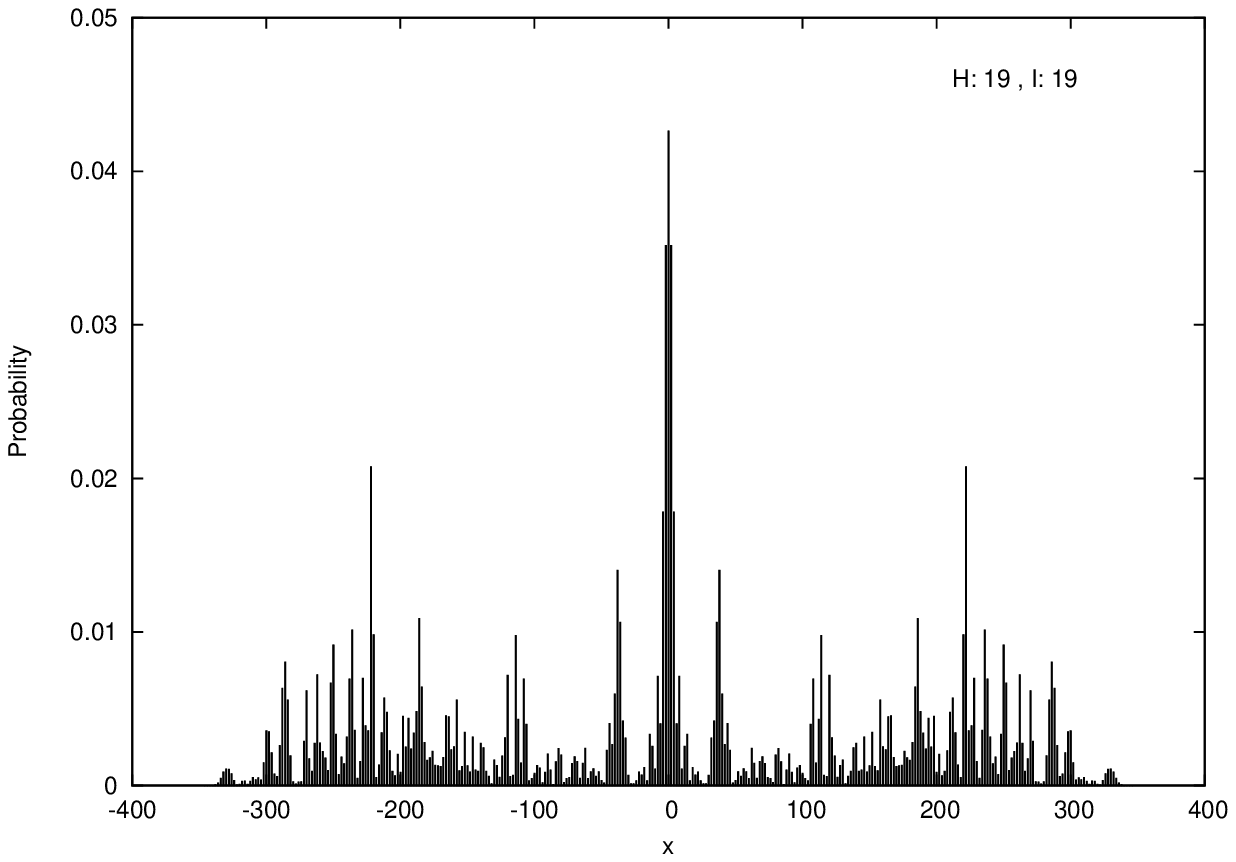}
\caption{ Probability $P(x,t)$ of the quantum walk of case IIIA ( $[H:19, I: 19]$)  with period $N=14$ at  $t=400$.  Localization at the origin occurs for $q=N/2 >13$.}\label{fig11}
\end{figure}

\begin{figure}
 \centering
 \includegraphics{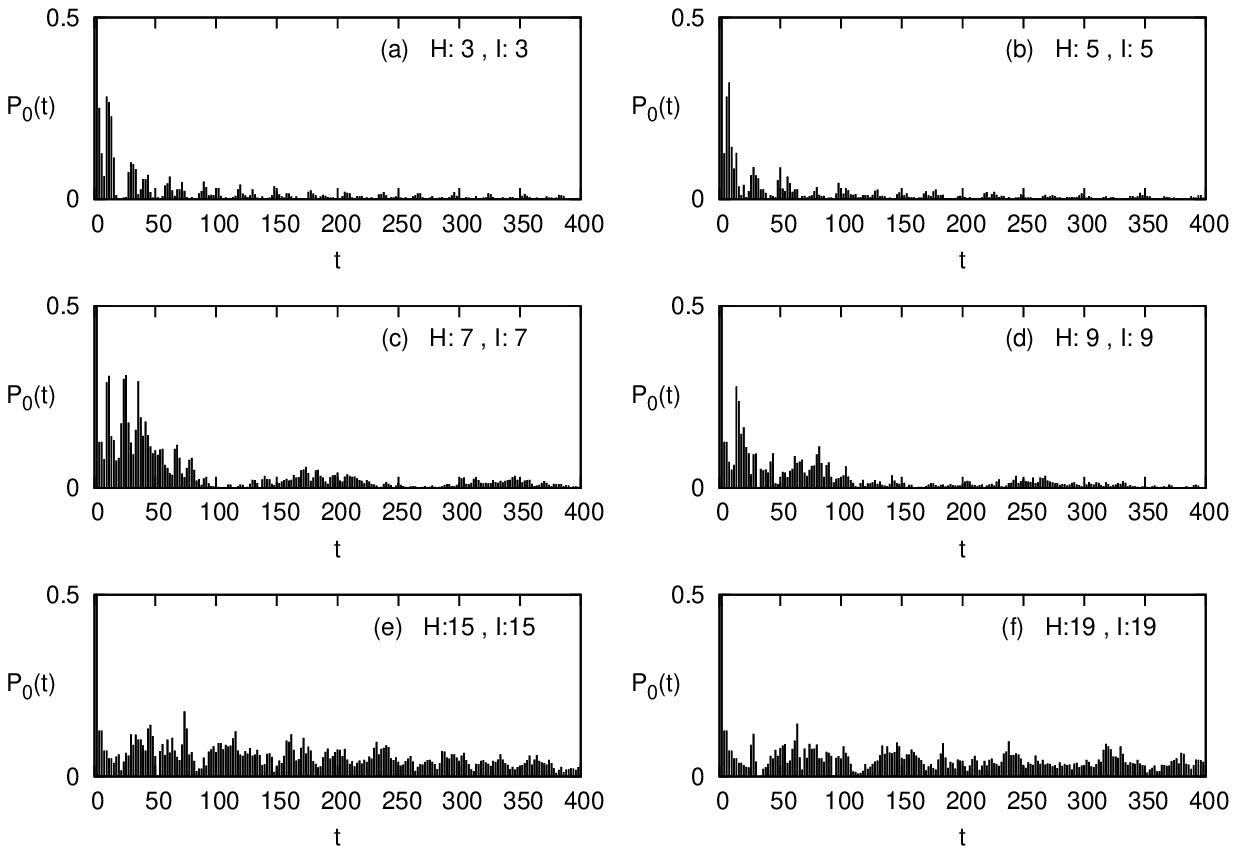}
\caption{Probability $P_0(t)$ at the origin of the quantum walk of  case IIIA.
The number of steps is taken up to 400. Significant recurrence occurs for $q>13$.}\label{fig12}
\end{figure}

For case IIIA localization effect is not as strong as the previous cases. 
Localization occurs only  for $q>13$.  Figs. \ref{fig11} and \ref{fig12}  show some representative situations.

\begin{figure}
 \centering
 \includegraphics{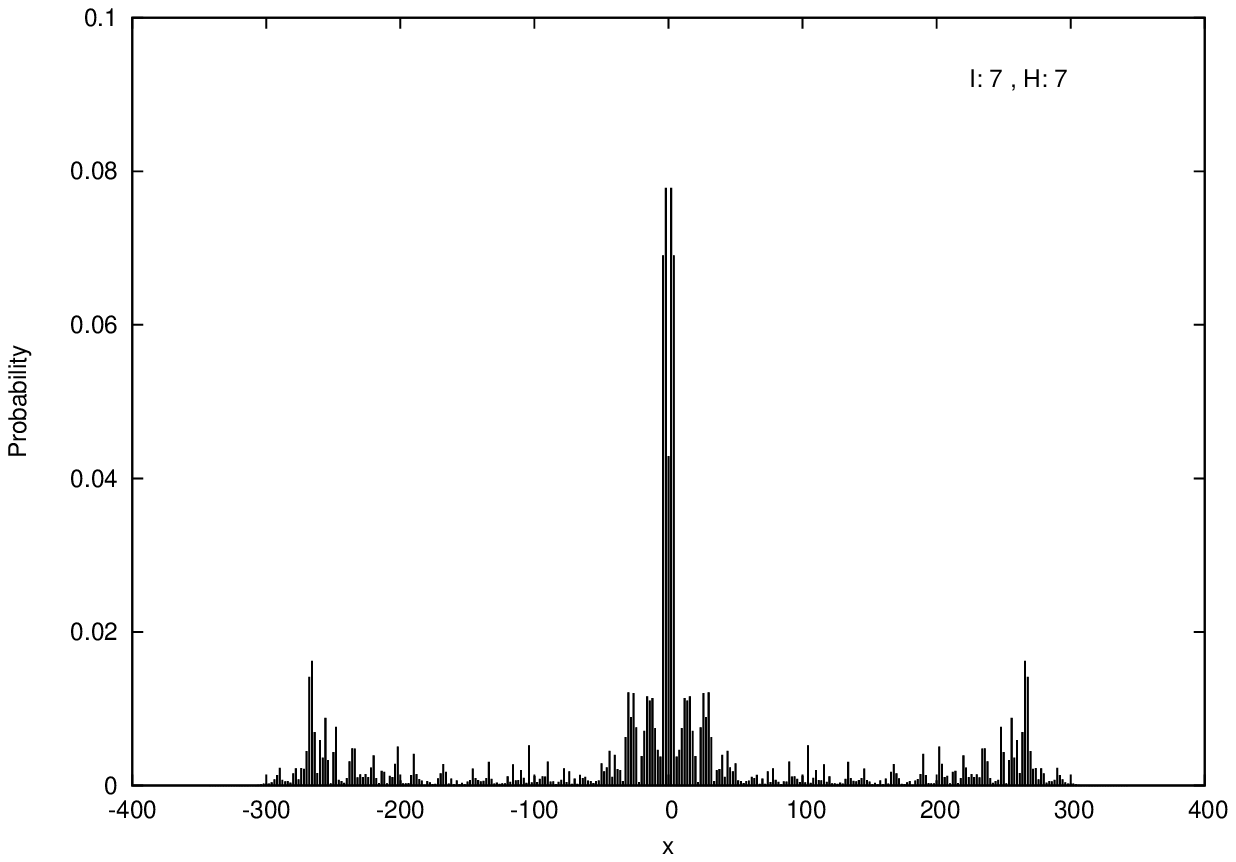}
\caption{ Probability $P(x,t)$ of the quantum walk of case IIIB ( $[I:7, H: 7]$)  with period $N=14$ at  $t=400$. Localization at the origin occurs for $q=N/2>3$.}\label{fig13}
\end{figure}

\begin{figure}
 \centering
 \includegraphics{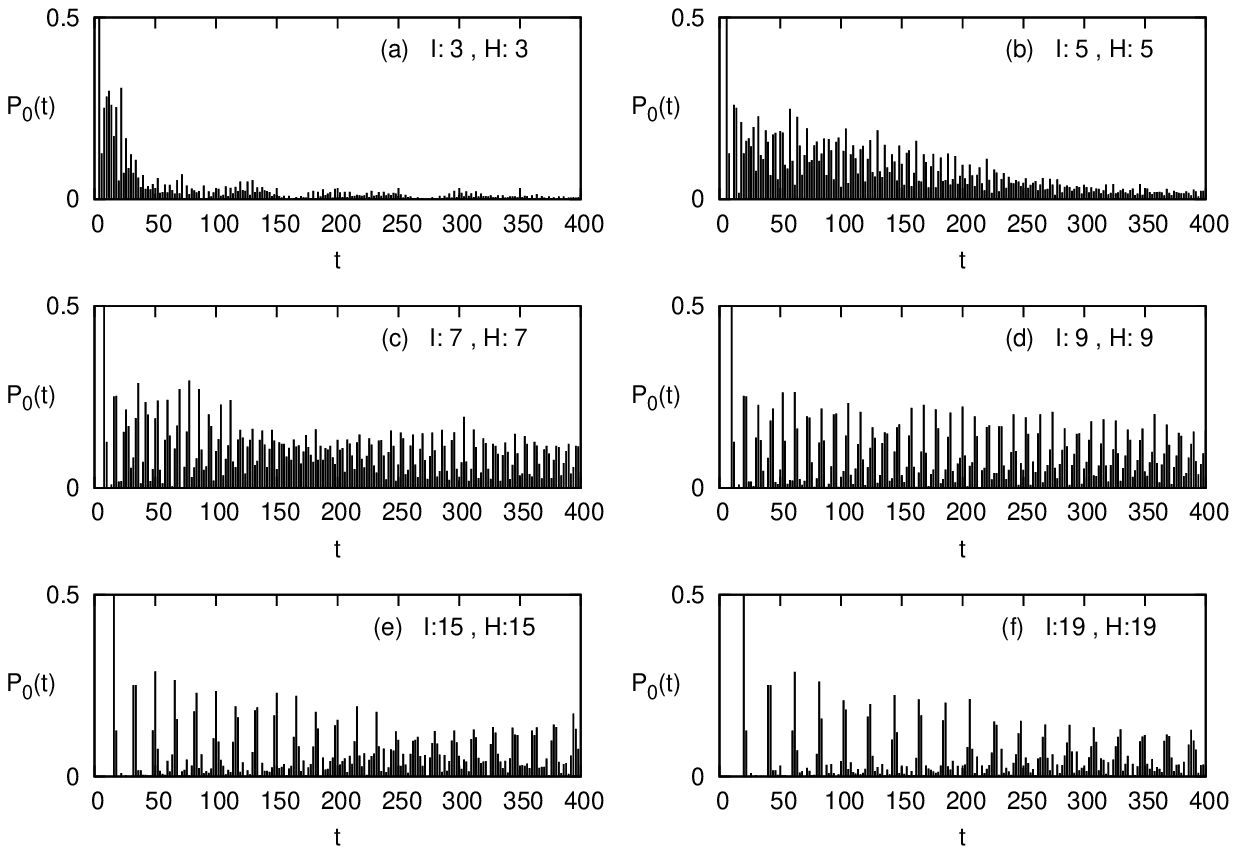}
 \caption{Probability $P_0(t)$ at the origin of the quantum walk of  case IIIB.
The number of steps is taken up to 400. Significant recurrence occurs for $q>3$. }
 \label{fig14}
 \end{figure}

Localization and recurrence occur  more  significantly for case IIIB. In fact, 
localization occurs already for $q>3$.  Figs.~\ref{fig13} and \ref{fig14} show these very clearly.  In fact,  of the six cases considered in this work, case IIIB shows the strongest effect of localization and recurrence.  Thus the tendency of diffusion is much slower in this case than in other cases.  This is also reflected in the fact, as depicted in Fig.~\ref{fig1}, that  the standard deviation $\sigma (t)$ of this case has the smallest slope than those of the other cases.

%
%

\section{Discussion}

We have presented a model of quantum walk in periodic potential on the line by taking the simple view that different potentials affect differently the way the coin state of the walker is changed.   For simplicity and definiteness, we assumed the walker's coin state is unaffected at sites without potential, and is rotated in an unbiased way according to
Hadamard matrix at sites with potential. 
This is the simplest and most natural model of a quantum walk in a periodic potential with two coins.

Six generic cases of such quantum walks were studied numerically.  It is found that of the six cases, two cases, case IA and IIB, behave in
a similar pattern as the original Hadamard walk where only a Hadamard coin is being used throughout the walk on the line. They show the 
same asymptotic values of the standard deviation as function of the step.  On the other hand, in the other four cases,   localization effect is possible, where the walker could be confined in the neighborhood of the origin for sufficiently long times. Associated with such localization effect is the recurrence of the probability of the walker returning to the neighborhood of the origin. 

A notable difference between the  cases IA and IIB with the other four cases is the number of the positions with potential, or the points where Hadamard coin is used within a single period of the potential:  it is smaller than the number of points without potential (or points using identity coin) .  Also, in the other four cases, localization and recurrence occur only when the number of points with Hadamard coin is ``sufficiently" larger than the number of points using identity coin.  This implies the existence of critical values of period $N$ for these cases.

Furthermore, it is also observed that the effect of  localization and recurrence occur more strongly in case IIIB and IB, as shown in Figs.~\ref{fig1} and \ref{fig2}. In these two cases, the walker starts at the center of the valley (with identity coin) of the potential. 

To summarize, it appears that in the model of quantum walk in periodic potential proposed in this work, localization and recurrence effect are stronger if the walker begin its walk in the middle of the valley (with identity coin) of a periodic  potential with a larger portion of potential (with Hadamard coin).

As a first attempt to study localization and recurrence phenomena of quantum walks in periodic potentials, we have relied on numerical method so far.   We hope to study this problem  with an analytic approach in the near future.  It is also interesting to study the behaviors of the quantum walk with other choices of the coins $C_0$ and $C_p$.   Generalization of the  present work to higher-dimensional cases is straightforward.   It would be nice if the quantum walks on periodic potential considered here could be experimentally implemented, say in optical lattice.


\section*{Acknowledgements}

This work is supported in part by the
Ministry of Science and Technology of Taiwan (R.O.C.)  under
Grants NSC-99-2112-M-032-002-MY3, NSC 102-2112-M-032-003-MY3
and in part by the National Center for Theoretical Sciences (North)  (NCTS-n) of R.O.C.
We would like to thank Y. Shikano for comments and for updating us on the relevant references.


\newpage

%
%

\end{document}